\documentclass[sigconf, nonacm]{acmart}
\pdfoutput=1

\usepackage{tikz}
\usetikzlibrary{shadows.blur}
\usepackage{pgfplots}
\usepgfplotslibrary{groupplots}
\usepackage{booktabs}
\usepackage{listings}

\usepackage{fancyvrb}
\usepackage{xcolor}

\newcommand\vldbdoi{XX.XX/XXX.XX}

\newcommand\vldbavailabilityurl{https://github.com/itrummer/CodexDB}
\begin{document}
%\title{NaturalDB: Synthesizing Code for Query Processing \\from Natural Language Instructions using GPT-3 [Vision]}
\title{CodexDB: Generating Code for Processing \\SQL Queries using GPT-3 Codex}

%\title{Synthesizing Code for Processing SQL Queries with GPT-3: \\First Results and a Research Road Map [Vision]}
%\title{Customizing Code Synthesis for Database Queries \\via Natural Language Hints Using GPT-3 [Vision]}

%%
%% The "author" command and its associated commands are used to define the authors and their affiliations.
\author{Immanuel Trummer}
\affiliation{%
  \institution{Cornell University}
  \city{Ithaca}
  \state{NY}
  \postcode{14850}
}
\email{itrummer@cornell.edu}

\begin{abstract}
CodexDB is an SQL processing engine whose internals can be customized via natural language instructions. CodexDB is based on OpenAI's GPT-3 Codex model which translates text into code. It is a framework on top of GPT-3 Codex that decomposes complex SQL queries into a series of simple processing steps, described in natural language. Processing steps are enriched with user-provided instructions and descriptions of database properties. Codex translates the resulting text into query processing code. An early prototype of CodexDB is able to generate correct code for a majority of queries of the WikiSQL benchmark and can be customized in various ways. 
\end{abstract}

\maketitle

% \pagestyle{\vldbpagestyle}
% \begingroup\small\noindent\raggedright\textbf{PVLDB Reference Format:}\\
% \vldbauthors. \vldbtitle. PVLDB, \vldbvolume(\vldbissue): \vldbpages, \vldbyear.\\
% \href{https://doi.org/\vldbdoi}{doi:\vldbdoi}
% \endgroup
% \begingroup
% \renewcommand\thefootnote{}\footnote{\noindent
% This work is licensed under the Creative Commons BY-NC-ND 4.0 International License. Visit \url{https://creativecommons.org/licenses/by-nc-nd/4.0/} to view a copy of this license. For any use beyond those covered by this license, obtain permission by emailing \href{mailto:info@vldb.org}{info@vldb.org}. Copyright is held by the owner/author(s). Publication rights licensed to the VLDB Endowment. \\
% \raggedright Proceedings of the VLDB Endowment, Vol. \vldbvolume, No. \vldbissue\ %
% ISSN 2150-8097. \\
% \href{https://doi.org/\vldbdoi}{doi:\vldbdoi} \\
% }\addtocounter{footnote}{-1}\endgroup

\ifdefempty{\vldbavailabilityurl}{}{
\vspace{.3cm}
\begingroup\small\noindent\raggedright\textbf{Artifact Availability:}\\
The source code, data, and/or other artifacts have been made available at \url{\vldbavailabilityurl}.
\endgroup
}

\section{Introduction}
\label{sec:intro}

Modifying a database management system is hard. Systems such as Postgres feature millions of code lines. Understanding and changing that code requires expert knowledge in databases on top of advanced coding skills. This prevents all but the most experienced developers from creating customized versions.

This paper presents the vision behind CodexDB, a novel database management system that can be customized without expert developer skills. Users specify natural language instructions, along with their queries, which influences code generated for query processing. The enabling technology for this system is OpenAI's GPT-3 Codex model. Codex is a large neural network, currently available via a private beta test, that translates natural language instructions into code. This paper presents first experimental results and an outlook on future steps.

The range of applications is vast. To name just a few, consider the following use cases.

%A developer wants to benchmark different Python frameworks (e.g., Pandas or Vaex) for processing a specific SQL workload on a specific hardware platforms. Traditionally, doing so requires either modifying an existing database management system or writing query-specific code from scratch. With CodexDB, that developer specifies queries, together with natural language instruction such as ``Use pandas library''. While CodexDB may not succeed at generating code for each workload query, obtaining performance results for a subset can guide future development efforts. Also, generated code can be manually validated and reused in case of recurrent query templates.

\begin{example}\label{ex:expert}
A developer wants to benchmark different data processing frameworks (e.g., Pandas and Vaex in Python or Tablesaw and Morpheus in Java) on a specific SQL workload and hardware platform. Traditionally, doing so requires either modifying an existing database management system or writing query-specific code from scratch. With CodexDB, that developer specifies queries, together with natural language instruction such as ``Use pandas library''. While CodexDB may not succeed at generating code for each workload query, obtaining performance results for a subset can guide future development efforts. Also, generated code can be manually validated and reused in case of recurrent queries.
\end{example}

\begin{example}\label{ex:novice}
A novice database user wants to gain a deeper understanding of how database management systems work. To that purpose, the user would like to generate customized output after each processing step (e.g., summarizing steps performed and showing a small sample of intermediate results). Integrating such changes into traditional systems is beyond the user's capabilities. With CodexDB, the user specifies natural language queries (which, internally, are translated into SQL), together with a natural language description of desired per-step output.
\end{example}

CodexDB accepts queries, together with natural language instructions, as input. These instructions customize the way in which queries are executed. CodexDB generates code to process queries while complying with additional instructions. A first option is to submit queries and instructions directly to GPT-3 for code generation. We will see in Section~\ref{sec:experiments} that this approach does not work. 

Instead, CodexDB adapts techniques from classical query planning. It decomposes complex SQL queries into sequences of simple processing steps. In contrast to prior work, those steps are formulated in natural language using corresponding text templates. Finally, automatically generated plan steps are interleaved with user-provided instructions. The resulting text is enriched with information about the database schema and physical layout. The final text is submitted to GPT-3 Codex (as a so-called ``prompt''). Using this approach as a starting point, CodexDB generates code for sample queries in a training step. The resulting code samples can be integrated into prompts generated at run time to increase the chances of success. An early prototype of CodexDB generates correct code in a majority of cases for a popular text-to-SQL benchmark. Also, it is able to customize generated code using simple instructions, inspired by the use cases outlined before.

In summary, the original scientific contributions in this paper are the following:

\begin{itemize}
    \item The paper presents the vision behind CodexDB, an analytical SQL engine that can be customized via natural language instructions.
    \item The paper discusses first experimental results, based on an early prototype of CodexDB.
    \item The paper outlines next steps and future research.% challenges
\end{itemize}

The remainder of this paper is organized as follows. Section~\ref{sec:related} discusses recent progress in natural language processing and compares CodexDB to prior work. Section~\ref{sec:overview} describes the architecture of the first prototype. Section~\ref{sec:experiments} reports first experimental results in multiple scenarios. Section~\ref{sec:future} discusses next steps and concludes.
\section{Background and Related Work}
\label{sec:related}

CodexDB is enabled by recent advances in the domain of natural language processing. Those advances have been fuelled by two key ideas: a novel neural network architecture, the Transformer~\cite{Vaswani2017}, and new training paradigms, implementing the idea of transfer learning~\cite{ruder2019transfer}. The Transformer is nowadays the dominant architecture in the domain of language processing~\cite{Wolf2020}. Among other advantages, it lends itself better to parallelization than prior methods. This has, in part, enabled the creation of very large, pre-trained language models. Such models are pre-trained on tasks for which large amounts of training data are easily available, e.g.\ predicting the next word in text snippets. While pre-training is very expensive, the resulting models can be easily specialized for new tasks via different methods. \textit{Fine-tuning} describes a process in which pre-trained models are used as a starting point for further training on more specialized tasks (reducing the amount of training samples and computational overheads by orders of magnitude via pre-training~\cite{Houlsby2019}). Until recently, fine-tuning has been the primary method of exploiting pre-trained language models. The latest generation of pre-trained models, most notably OpenAI's Generative Pre-Trained Transformer (GPT) version 3, unlocks new possibilities. It turns out that sufficiently large models can oftentimes solve new tasks without specialized training (``zero-shot learning''), based on inputs describing the task in natural language alone~\cite{Brown2020}. Precision increases if the input integrates few (i.e., typically less than ten) examples pairing tasks of the same type with solutions (``few-shot learning''). This is the method currently used by CodexDB. The final development that enabled this paper is the emergence of the Codex variant of GPT-3~\cite{Chen2021, OpenAI2021}. The primary difference between GPT-3 Codex and the original GPT-3 model lies in the data used for pre-training. GPT-3 Codex is trained on code and technical documentation. This results in a model whose primary use case is the translation of natural language commands into code.

CodexDB connects to prior work on natural language interfaces in the database community~\cite{Li2014, Saha2016, Weir2019}. So far, the focus was on ``democratizing access to data'', i.e.\ enabling lay users to work with database systems. CodexDB goes one step further by ``democratizing'' the design of database system internals. The goal is to enable lay users to change system behavior to a degree that goes beyond the configuration scope of traditional database systems (as well as making such changes easier for more advanced users). 

CodexDB relates to prior work exploiting machine learning~\cite{Kipf2018, Kraska2017, Marcus2018a} and specifically Transformers~\cite{Suri2021, Tang2021} in the context of database systems. It connects broadly to prior work using GPT-3 for program synthesis~\cite{Jain2021, Liguori2022, Li2022}. It differs by its focus on customizable SQL query processing. Prior work on code generation for query processing~\cite{Krikellas2010a, Wanderman-Milne2014} cannot integrate natural language instructions. 

%More broadly, CodexDB connects to other work exploiting similar technologies~\cite{Suri2021, Tang2021}, in particular recent work based on GPT-3 Codex~\cite{Jain2021, Liguori2022, Li2022}. CodexDB differs by its focus on customizable SQL query processing.

\section{System Overview}
\label{sec:overview}

\tikzstyle{system}=[draw, rectangle, rounded corners=1mm, shade, top color=gray!10, bottom color=gray!20, blur shadow={shadow blur steps=5}]
\tikzstyle{component}=[draw, rectangle, shade, top color=blue!10, bottom color=blue!20, minimum width=4cm, font=\bfseries, blur shadow={shadow blur steps=5}]
\tikzstyle{dataflow}=[thick, -latex]
\tikzstyle{mdataflow}=[thick, <->]
\tikzstyle{io}=[font=\itshape]

\begin{figure}[t]
    \centering
    \begin{tikzpicture}
        \draw[system] (-2.25,-5.5) rectangle (3.5,1);
        \node[font=\bfseries\Large] at (0.625,0.7) {CodexDB};
        \node[component] (nlqi) at (0,0) {Text-to-SQL (Optional)};
        \node[component] (planner) at (0,-1) {Natural Language Planner};
        \node[io, anchor=east] (query) at (-2.5,0) {Query};
        \node[io, anchor=east] (mod) at (-2.5,-1) {Instructions};
        \node[component] (prompt) at (0,-2) {Prompt Generator};
        \node[component] (code) at (0,-3) {Code Generator (GPT-3)};
        \node[component] (engine) at (0,-4) {Execution Engine};
        \node[component] (verify) at (0,-5) {Verification};
        \node[component, rotate=90, minimum width=2.5cm] (samples) at (3,-4) {Code Samples};
        \node[component, rotate=90, minimum width=2.5cm] (catalog) at (3,-1) {DB Catalog};
        \node[io, anchor=east] (result) at (-2.5,-4) {Result};
        
        \draw[dataflow] (nlqi) -- (planner);
        \draw[dataflow] (planner) -- (prompt);
        \draw[dataflow] (prompt) -- (code);
        \draw[dataflow] (code) -- (engine);
        \draw[dataflow] (engine) -- (verify);
        \draw[dataflow] (verify.east) -- (samples);
        \draw[dataflow] (catalog) -- (prompt);
        \draw[dataflow] (samples.east) -- (prompt);
        \draw[dataflow] (query) -- (nlqi);
        \draw[dataflow] (mod) -- (planner);
        \draw[dataflow] (engine) -- (result);
    \end{tikzpicture}
    \caption{Overview of CodexDB prototype.}
    \label{fig:overview}
\end{figure}
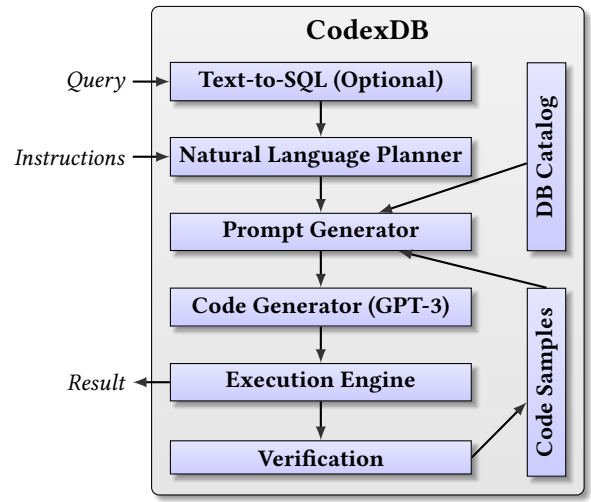

Figure~\ref{fig:overview} shows an overview of CodexDB. Users enter a query as well as natural language instructions, influencing the code generated for query processing. The query is formulated either in SQL or in natural language. In the latter case, the query is first translated into an SQL query via text-to-SQL methods~\cite{Arbor2016, Sen2020, Zhong2017}. 

\begin{table}[t]
    \centering
    \caption{Text templates used during planning.}
    \begin{tabular}{ll}
         \textbf{Pattern} & \textbf{Text Template} \\
         \midrule[1pt]
         \verb|X = Y| & Check if T(\verb|X|) equals T(\verb|Y|) \\
         \midrule
         \verb|from X where Y| & Filter T(\verb|X|) using T(\verb|Y|) \\
         \midrule
         \verb|X as Y| & T(\verb|X|) (aka.\ T(\verb|Y|)) \\
         \midrule
         \verb|select X from Y| & Create table with columns T(\verb|X|) from T(\verb|Y|) \\
         \bottomrule[1pt]
    \end{tabular}
    \label{tab:templates}
\end{table}

The SQL query and natural language instructions form the input to the query planner. This planner differs from prior query planners by its output format. As the plan is translated into code by GPT-3 in the following steps, the plan is formulated as a sequence of natural language steps. User-provided natural language instructions are included as steps in such plans. 

More precisely, the planner treats the nodes in the query tree in post-order. Each node is translated into a processing step, formulated in natural language. To do so, the planner uses text templates that are associated with specific node types. Table~\ref{tab:templates} shows example templates (T(\verb|X|) and T(\verb|Y|) denote the text representation of expressions \verb|X| and \verb|Y| respectively). As plan steps are numbered, intermediate results are referred to via the number of the step generating them. The last step in the plan instructs GPT-3 to write the query result into a file at a specified location. 

Currently, CodexDB allows users to specify two types of instructions: instructions that refer to the plan execution as a whole (e.g., instructions on which libraries to use for processing) as well as instructions that are executed after each step (e.g., instructions determining customized logging output). Instructions of the former type are pre-pended to the template-based processing steps (i.e., they become the first plan step) while instructions of the latter type are inserted after each processing step (i.e., the number of plan steps doubles). Note that, currently, the planner does not perform any cost-based or heuristic optimization.

Code generation is initiated by submitting a prompt to GPT-3 for completion. This prompt represents the start of a program that GPT-3 Codex tries to finish. The prompt integrates details about the data in the database, extracted from the database catalog, including the names of tables and their columns, as well as a path to the corresponding files. This description is generated using a simple text template with placeholders for column and table names. Also, the prompt integrates the aforementioned plan steps. The prompt is passed on to GPT-3 which answers with a piece of code. CodexDB tries to execute the code, and to read the generated query result. If the code does not execute (or if it does not generate a result file), CodexDB executes up to a configurable number of retries. With each retry, the ``temperature'' (a Codex parameter determining the degree of randomness in code generation) is increased to enable new solutions. If successful, the result is returned to the user. 

CodexDB can be used in a zero-shot setting (i.e., it generates code with instructions not seen before). Alternatively, it executes a training phase before run time with fixed instructions. The purpose of training is to generate a library of code samples, generated using the target instructions. During training, CodexDB uses sample queries for which the query result is known. It retries code generation until the query result matches the known one (or until it reaches the maximal number of retries). At run time, a specified number of samples is randomly selected from that library and included into the prompt. Having examples in the prompt (``few-shot learning'') increases the success probability, as shown in Section~\ref{sec:experiments}.

%If generating code samples for the library, before run time, CodexDB may verify the query result by comparing to a known ground truth result. If the ground truth is known, CodexDB may retry until the query result matches the reference. 

%Optionally, the prompt may contain examples of completed programs. If enabled, CodexDB integrates a specified number of examples, selected randomly from a library of samples, into the prompt. Integrating examples increases the success rate of code generation (see Section~\ref{sec:experiments}).

%The description follows a simple text template with placeholders for column and table names. 
%Currently, all database files are stored in the .csv format. 

%The prompt is passed on to GPT-3 which answers with a piece of code. CodexDB writes that code into a file on hard disk, executes the code, and tries to extract the query result from a file on hard disk. If the code does not execute (or if it does not generate a result file), CodexDB executes up to a configurable number of retries. If successful, the result is returned to the user. 

%If generating code samples for the library, before run time, CodexDB may verify the query result by comparing to a known ground truth result. If the ground truth is known, CodexDB may retry until the query result matches the reference. 

\begin{figure}
\small
    \centering
\begin{Verbatim}[commandchars=\\\{\}]
"""
\textcolor{blue}{Table Data with columns 'Player','No_','Nationality','Position',}
\textcolor{blue}{'Years_in_Toronto','School_Club_Team', stored in 'Data.csv'.}
Processing steps:
1. Load data for table Data.
\textcolor{red}{2. Print progress updates.}
3. Check if 'Player' equals 'dell curry'.
\textcolor{red}{4. Print progress updates.}
5. Filter results of Step 1 using results of Step 3.
\textcolor{red}{6. Print progress updates.}
7. Create table with columns 'Years_in_Toronto' 
   (aka. result ) from results of Step 5.
\textcolor{red}{8. Print progress updates.}
9. Write results of Step 7 to file 'result.csv' (with header).
\textcolor{red}{10. Print progress updates.}
"""
\end{Verbatim}
    \caption{Example prompt for code generation integrating a description of the database (blue), processing steps (black), and natural language instructions (red).}
    \label{fig:exampleprompt}
\end{figure}

\lstset{showspaces=false, showstringspaces=false}

\begin{figure}
    \centering
    \small
\begin{lstlisting}[language=Python]
data = pd.read_csv('Data.csv')
print('Loaded Data')
is_dell_curry = data['Player'] == 'dell curry'
print('Checked if Player equals dell curry')
dell_curry_data = data[is_dell_curry]
print('Filtered data for dell curry')
result = dell_curry_data['Nationality']
print('Created table with Nationality column')
result.to_csv('result.csv', header=True)
print('Wrote to file result.csv')
\end{lstlisting}
    
    \caption{Code generated by GPT-3 in response to prompt from Figure~\ref{fig:exampleprompt} (code was shortened for readability by removing empty lines and comments).}
    \label{fig:examplegenerated}
\end{figure}

\begin{example}
Consider the query \verb|Select "Years_in_Toronto"| \verb|as Result from Data where "Player" = 'dell curry'| from the WikiSQL benchmark. Assume a user enters this query, together with the per-step instructions ``Print progress updates''. Figure~\ref{fig:exampleprompt} shows the prompt for this query, interleaving automatically generated processing steps with user instructions and providing context on the database schema. Figure~\ref{fig:examplegenerated} shows the generated code. It processes the query and writes the result into a file. While doing so, it prints out progress updates summarizing steps performed.
\end{example}
\section{Experiments}
\label{sec:experiments}

\pgfplotscreateplotcyclelist{linelist}{{blue, solid, mark=*},{red, mark=x, mark size=3},{brown, mark=diamond, mark size=3},{gray, mark=triangle, mark size=3},{purple, mark=+, mark size=5},{violet, mark=square, dotted, thick, mark size=3},{cyan, mark=*, mark size=2}}

The goal of the experiments is threefold. First, to verify that CodexDB generates correct code in most cases. Second, to evaluate the degree to which code can be customized via natural language instructions. Third, to compare CodexDB to other baselines. Section~\ref{sub:setup} discusses the experimental setup while Section~\ref{sub:results} reports results.

\subsection{Setup}
\label{sub:setup}

All experiments are executed on an AWS EC2 instance of type t2.xlarge with 16~GB of RAM, four virtual CPUs, and 800~GB of EBS storage. The instance uses Amazon's Deep Learning AMI (Version~53) and runs Ubuntu~18.04. CodexDB is implemented in Python~3 and accesses OpenAI's GPT-3 Codex model via OpenAI's Python API. The experiments use the ``Cushman'' and ``Davinci'' versions of Codex with an estimated parameter count of 6.7~billion and 175~billion parameters respectively~\cite{Brown2020, textToSQLCodex21}. The generated code is in Python, the language both models are most capable in~\cite{OpenAI2021}.

The following experiments compare CodexDB to baselines that try translating natural language queries directly to code. This is the most direct method of using GPT-3, making the comparison interesting. Doing so requires a text-to-SQL benchmark that features natural language questions, along with corresponding queries. We consider a subset of the WikiSQL benchmark~\cite{Zhong2017}, a popular benchmark featuring over 80,000 queries with examples. The experiments only consider up to the first hundred queries as treating all queries is prohibitively expensive\footnote{At the time of writing, OpenAI Codex is only available to beta testers and access is subject to a rate limit of 20 requests per minute.}. The data on which queries operate is stored in the .csv format.

The experiments evaluating CodexDB focus on the key step of translating an SQL query into code, possibly with additional natural language instructions. Translating natural language questions into SQL queries is a well studied problem. Corresponding results for the WikiSQL benchmark are available~\cite{Zhong2017} with recent methods achieving a precision of over 90~\%~\cite{Xuan2021}. We consider a test case (characterized by a natural language query with associated data) as ``solved'' if the generated program is executable and generates the correct result. This proxy for correctness is often used to evaluate natural language query interfaces~\cite{Zhong2017, Scholak2021}. A subset of generated programs was manually validated as well. Unless noted otherwise, CodexDB retries generating a program once if the first generated program is not executable. If the first program executes but generates an incorrect result, the corresponding test case is not solved. CodexDB uses a temperature of zero for the first try and increases the temperature (determining the degree of randomization during code generation) by an amount determined by the formula $0.5/N$ where $N$ is the maximal number of allowed tries (typically two).

To test customization, we consider six natural language instructions. Three of them focus on processing methods by instructing CodexDB to use specific libraries: ``Use pandas library'', ``Use vaex library'', and ``Use datatable library''. The other three instruct CodexDB to generate specific logging output after each processing step: ``Print 'Done.' '', ``Print intermediate results'', and ``Print progress updates''. The first three instructions are added once as first plan step. The last three are added after each step of the initial plan. Note that the following figures and tables abbreviate those instructions slightly (e.g., in the figure legends).

\subsection{Results}
\label{sub:results}

\begin{figure}[t]
    \centering
        \begin{tikzpicture}
        \begin{groupplot}[group style={group size=1 by 1, ylabels at=edge left, xlabels at=edge bottom}, width=6cm, height=3.5cm, legend entries={Question Prompt, Query Prompt, CodexDB Prompt}, legend columns=1, ymode=normal, ymajorgrids, xlabel={GPT-3 Codex variant}, ylabel={Nr.\ Solved}, legend pos=outer north east, ylabel near ticks, xlabel near ticks, xtick={0, 1}, xticklabels={Cushman, Davinci}, ybar, nodes near coords, enlargelimits=0.275, xmin=-0.5, xmax=1.5]
        \nextgroupplot
        \addplot coordinates {(0, 0) (1, 0)};
        \addplot coordinates {(0, 0) (1, 0)};
        \addplot coordinates {(0, 11) (1, 22)};
        \end{groupplot}    
    \end{tikzpicture}
    
    \caption{Number of test cases solved out of 100 without prior training (``zero-shot'') for different models and prompts.}
    \label{fig:zero}
\end{figure}
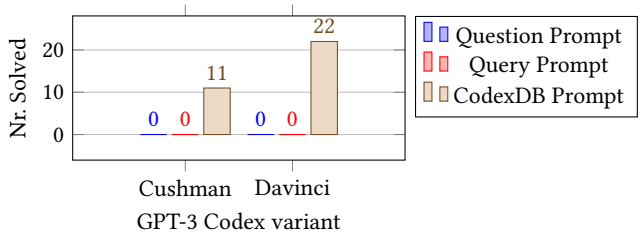

Figure~\ref{fig:zero} reports results of an experiment comparing different prompt generation methods (on 100 queries from the WikiSQL test set). ``CodexDB Prompt'' refers to prompts generated by CodexDB (integrating, in particular, natural language query plans). ``Question Prompt'' and ``Query Prompt'' integrate the same description of the data source as CodexDB (i.e., table and column names) but replace the natural language query plan by the natural language question or the correct SQL query respectively. Clearly, the prompts of CodexDB, enriched by query plans, are necessary to generate correct code. The Davinci model (which features most parameters) solves significantly more test cases than the Cushman version. On the other side, average generation times (seven seconds versus two seconds) are higher for Davinci.

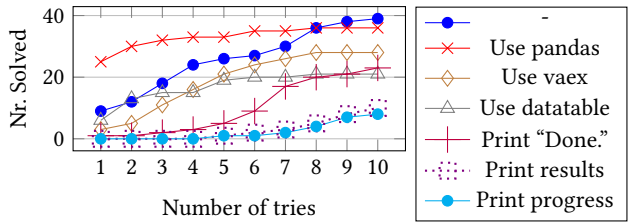
\begin{figure}
    \centering
    \begin{tikzpicture}
        \begin{axis}[xlabel={Number of tries}, ylabel={Nr.\ Solved}, width=6cm, height=3.5cm, ymajorgrids, xtick=data, ylabel near ticks, legend entries={-,Use pandas,Use vaex,Use datatable,Print ``Done.'',Print results,Print progress}, legend columns=1, legend pos=outer north east, cycle list name={linelist}]
            \addplot coordinates {(1, 9) (2, 12) (3, 18) (4, 24) (5, 26) (6, 27) (7, 30) (8, 36) (9, 38) (10, 39)};
            \addplot coordinates {(1, 25) (2, 30) (3, 32) (4, 33) (5, 33) (6, 35) (7, 35) (8, 36) (9, 36) (10, 36)};
            \addplot coordinates {(1, 3) (2, 5) (3, 11) (4, 16) (5, 21) (6, 24) (7, 26) (8, 28) (9, 28) (10, 28)};
            \addplot coordinates {(1, 6) (2, 13) (3, 15) (4, 15) (5, 19) (6, 20) (7, 20) (8, 21) (9, 21) (10, 21)};
            \addplot coordinates {(1, 1) (2, 1) (3, 2) (4, 3) (5, 5) (6, 9) (7, 17) (8, 20) (9, 21) (10, 23)};
            \addplot coordinates {(1, 0) (2, 0) (3, 0) (4, 0) (5, 1) (6, 2) (7, 3) (8, 5) (9, 8) (10, 10)};
            \addplot coordinates {(1, 0) (2, 0) (3, 0) (4, 0) (5, 1) (6, 1) (7, 2) (8, 4) (9, 7) (10, 8)};
        \end{axis}
    \end{tikzpicture}
    \caption{Number of test cases out of 50 solved during training for different instructions as function of the number of tries.}
    \label{fig:training}
\end{figure}

Figure~\ref{fig:zero} reports a success rate of 22\% without prior training. Language models are often fine-tuned to increase performance for specific tasks. This option is not yet available for the Codex series of GPT-3. Instead, we consider few-shot scenarios~\cite{Brown2020} in the following. Here, examples with solutions are integrated as part of the prompt. 

\begin{figure}[t]
    \centering
        \begin{tikzpicture}
        \begin{groupplot}[group style={group size=2 by 1, y descriptions at=edge left, xlabels at=edge bottom, horizontal sep=5pt}, width=3.5cm, height=3.5cm, ymode=normal, ymajorgrids, yminorgrids, ytick={0, 20, 40, 60, 80, 100}, xlabel={Nr.\ samples}, ylabel={Nr.\ Solved}, ylabel near ticks, xlabel near ticks, xtick={0, 2, 4}, enlargelimits=0.1, ymin=0, ymax=100, cycle list name={linelist}]
        \nextgroupplot[title=Cushman Model]
        \addplot coordinates {(0, 0) (2, 59) (4, 63)};
        \addplot coordinates {(0, 0) (2, 62) (4, 54)};
        \addplot coordinates {(0, 11) (2, 67) (4, 23)};
            
        \nextgroupplot[title=Davinci Model,legend entries={Question Prompt, Query Prompt, CodexDB Prompt}, legend columns=1, legend pos=outer north east]
        \addplot coordinates {(0, 0) (2, 68) (4, 65)};
        \addplot coordinates {(0, 0) (2, 58) (4, 61)};
        \addplot coordinates {(0, 22) (2, 79) (4, 73)};
        \end{groupplot}    
    \end{tikzpicture}
    \caption{Number of test cases solved out of 100 as a function of the number of training samples in prompt (``few-shot'').}
    \label{fig:plain}
\end{figure}
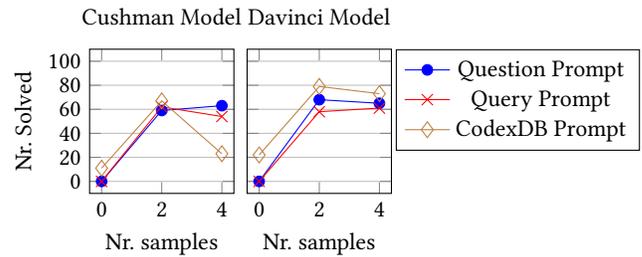

Figure~\ref{fig:training} reports the results of a preparation run, using 50~queries from the WikiSQL training set and the Davinci model. As training is executed before run time, up to ten tries are allowed. Furthermore, it is assumed that solutions for training samples are available, allowing to stop code generation only if the execution result is correct (as opposed to using the first executable code). Figure~\ref{fig:plain} reports solved test cases as a function of the (maximal) number of tries. Different lines are associated with additional natural language instructions (``-'' designates no additional instructions). Training took between 1510 seconds (when instructed to use the pandas library) and 8,300 seconds (with instructions ``print intermediate results''). Given enough tries and results to compare to, CodexDB solves 80\% of test cases without additional instructions.

Figure~\ref{fig:plain} reports number of test cases solved (out of 100 queries from the WikiSQL \textit{test set}, i.e.\ no overlap with pre-generated samples) as a function of the number of samples included in the prompt. It compares the previously introduced prompt styles. Clearly, performance improves significantly (e.g., from around 20 to around 80\% for Davinci) when adding samples. Adding samples decreases the gap between CodexDB's and other prompts. Still, the CodexDB prompt performs best except for four samples and the Cushman model. The reason is the slightly longer prompts of CodexDB (featuring query plans) that exceed the maximum input size for the Cushman model for 67 test cases. Davinci supports larger inputs and does not suffer from this problem. Unless noted otherwise, the remaining experiments use two samples and the Davinci model (the configuration leading to maximal performance in Figure~\ref{fig:plain}).

%For Cushman and plan prompts, 67 test cases exceed the maximal prompt length when using four examples.

%Median generation times: 2 seconds for Cushman, 7 seconds for Davinci.

\begin{table}
    \centering
    \caption{Code length in characters for different languages and instructions (only considering executable programs).}
    \begin{tabular}{lp{3.25cm}lll}
    \toprule[1pt]
         Language & Instructions & Min & Median & Max \\
    \midrule[1pt]
         SQL & - & 42 & 77 & 227 \\
         \midrule
         Python & - & 276 & 545 & 1110 \\
         \midrule
         & Use pandas library & 284 & 438.5 & 782 \\
         \midrule
         & Use vaex library & 355 & 637 & 1018 \\
         \midrule
         & Use datatable library & 307 & 437 & 848 \\
         \midrule
         & Print ``Done.'' & 390 & 724 & 1388 \\
         \midrule
         & Print intermediate results & 458 & 875 & 1734 \\
         \midrule
         & Print progress updates & 585 & 836 & 1458 \\
         \bottomrule[1pt]
    \end{tabular}
    \label{tab:codesize}
\end{table}

\begin{figure}[t]
    \centering
        \begin{tikzpicture}
        \begin{groupplot}[group style={group size=1 by 2, ylabels at=edge left, xlabels at=edge bottom}, width=8cm, height=3.5cm, legend entries={-,Use pandas,Use vaex,Use datatable, Print ``Done.'', Print results, Print progress}, legend columns=3, ymode=normal, ymajorgrids, ylabel={Nr.\ Solved}, legend to name=modificationsLegend, ylabel near ticks, xlabel near ticks, xtick=\empty, ybar, nodes near coords, enlargelimits=0.275]
        \nextgroupplot[title=Cushman-Codex Model, bar width=7pt]
        \addplot coordinates {(2, 67)};
        \addplot coordinates {(2, 69)};
        \addplot coordinates {(2, 74)};
        \addplot coordinates {(2, 60)};
        \addplot coordinates {(2, 64)};
        \addplot coordinates {(2, 69)};
        \addplot[fill=black] coordinates {(2, 72)};

        \nextgroupplot[title=Davinci-Codex Model, bar width=7pt]
        \addplot coordinates {(2, 79)};
        \addplot coordinates {(2, 81)};
        \addplot coordinates {(2, 76)};
        \addplot coordinates {(2, 58)};
        \addplot coordinates {(2, 69)};
        \addplot coordinates {(2, 71)};
        \addplot[fill=black] coordinates {(2, 77)};
        \end{groupplot}
    \end{tikzpicture}
    
    \ref{modificationsLegend}
    \caption{Number of test cases solved out of 100 for different natural language instructions.}
    \label{fig:modifications}
\end{figure}
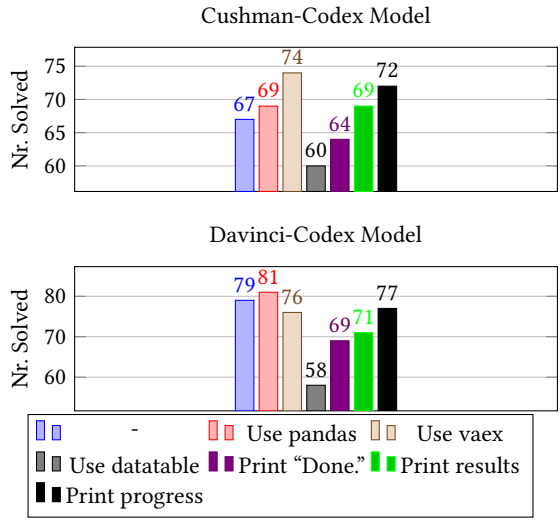

We test customization by adding the instructions described in Section~\ref{sub:setup}. Figure~\ref{fig:modifications} reports the number of test cases solved with different instructions. In most cases, adding more instructions tends to decrease success ratio for the largest model. Interestingly, the impact varies across instructions. In particular, asking CodexDB to use the pandas library slightly \textit{increases} performance. This seems reasonable as the pandas library is popular (i.e., the training set of GPT-3 Codex likely includes various example codes) and supports operations similar to SQL operators. Manual analysis of the first 20 programs generating the correct result shows that they are indeed correct. Table~\ref{tab:codesize} reports statistics on the size of generated code (and on the size of the corresponding SQL queries), measured in characters. The average size of code generated by CodexDB is larger by up to one order of magnitude, compared to SQL queries. This illustrates the difficulty of the task. Adding instructions on logging increases code size (due to print statements after processing steps).

\begin{figure}[t]
    \centering
        \begin{tikzpicture}
        \begin{groupplot}[group style={group size=1 by 1, ylabels at=edge left, xlabels at=edge bottom}, width=8cm, height=3.5cm, legend entries={-,Use pandas,Use vaex,Use datatable}, legend columns=4, ymode=normal, ymajorgrids, xlabel={Imported libraries}, ylabel={Nr.\ Programs}, legend to name=plainLegend, ylabel near ticks, xlabel near ticks, xtick=data, xticklabels={csv, pandas, vaex, datatable}, ybar, nodes near coords, enlargelimits=0.275]
        \nextgroupplot[title=Davinci-Codex Model, bar width=7pt]
        \addplot coordinates {(0,34) (1,66) (2,0) (3,0)};
        \addplot coordinates {(0,0) (1,100) (2,0) (3,0)};
        \addplot coordinates {(0,16) (1,61) (2,100) (3,0)};
        \addplot coordinates {(0,3) (1,15) (2,0) (3,100)};
        \end{groupplot}
    \end{tikzpicture}
    
    \ref{plainLegend}
    \caption{Number of generated programs out of 100 importing specific libraries for library-related instructions.}
    \label{fig:libraries}
\end{figure}
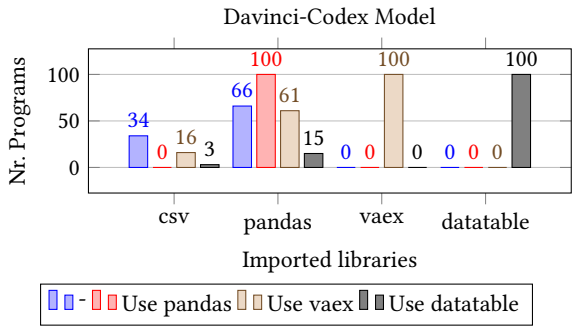

So far, we discussed correctness. Next, we examine whether additional instructions are reflected in the generated programs. Figure~\ref{fig:libraries} reports the number of generated programs (out of 100) that import certain libraries. Without specific instructions, 34\% of generated programs import the ``csv'' library while 66\% import pandas. Incorporating instructions to use pandas, vaex, or datatable into the prompt ensures that each generated programs imports the associated library. In some cases, in particular for vaex, programs import multiple libraries (both, csv and pandas). Manual inspection of the generated code reveals that some of those programs contain redundancy (e.g., by importing data using vaex, then transforming into pandas data frames). While this subset of programs formally satisfies the instructions (they import, i.e.\ ``use'', the corresponding library), they do not entirely reflect its spirit.

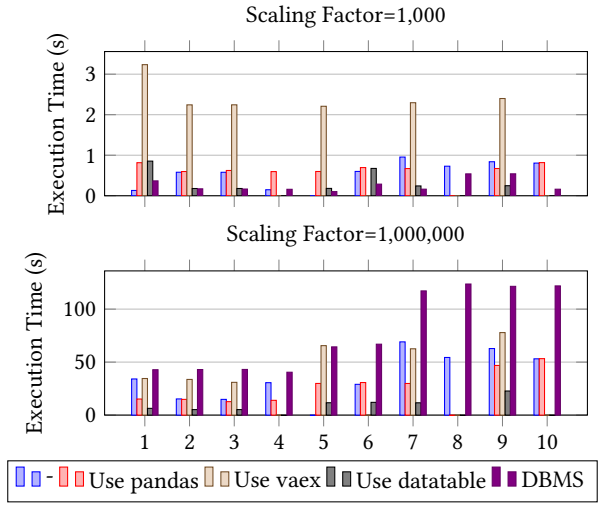
\begin{figure}[t]
    \centering
        \begin{tikzpicture}
        \begin{groupplot}[group style={group size=1 by 2, ylabels at=edge left, x descriptions at=edge bottom}, width=8cm, height=3.5cm, legend entries={-,Use pandas,Use vaex,Use datatable, DBMS}, legend columns=5, xtick=data, ymode=normal, ymajorgrids, ylabel={Execution Time (s)}, legend to name=scalingLegend, ylabel near ticks, xlabel near ticks, ybar=0pt, ymin=0]
        \nextgroupplot[title={Scaling Factor=1,000}, bar width=2pt]
        \addplot table[col sep=comma, x expr={\pgfplotstablerow+1}, y index=0, skip first n=1] {plots/performance/plain/times.csv};
        \addplot table[col sep=comma, x expr={\pgfplotstablerow+1}, y index=0, skip first n=1] {plots/performance/pandas/times.csv};
        \addplot table[col sep=comma, x expr={\pgfplotstablerow+1}, y index=0, skip first n=1] {plots/performance/vaex/times.csv};
        \addplot table[col sep=comma, x expr={\pgfplotstablerow+1}, y index=0, skip first n=1] {plots/performance/datatable/times.csv};
        \addplot table[col sep=comma, x expr={\pgfplotstablerow+1}, y index=0, skip first n=1] {plots/performance/sql/times.csv};

        \nextgroupplot[title={Scaling Factor=1,000,000}, bar width=2pt]
        \addplot table[col sep=comma, x expr={\pgfplotstablerow+1}, y index=1, skip first n=1] {plots/performance/plain/times.csv};
        \addplot table[col sep=comma, x expr={\pgfplotstablerow+1}, y index=1, skip first n=1] {plots/performance/pandas/times.csv};
        \addplot table[col sep=comma, x expr={\pgfplotstablerow+1}, y index=1, skip first n=1] {plots/performance/vaex/times.csv};
        \addplot table[col sep=comma, x expr={\pgfplotstablerow+1}, y index=1, skip first n=1] {plots/performance/datatable/times.csv};
        \addplot table[col sep=comma, x expr={\pgfplotstablerow+1}, y index=1, skip first n=1] {plots/performance/sql/times.csv};
        \end{groupplot}
    \end{tikzpicture}
    
    \ref{scalingLegend}
    \caption{Execution time of programs generated by CodexDB with different instructions and of one traditional DBMS.}
    \label{fig:scaling}
\end{figure}

%Scaled up data sets with highest scaling factor: 1.2~GB mean size, 2.2~GB max size. 27~M rows maximum, 15~M is the mean.

% in Figure~\ref{fig:scaling}

\begin{table}
    \centering
    \caption{Total run time for queries solved by all baselines.}
    \begin{tabular}{lrr}
    \toprule[1pt]
         \textbf{Baseline} & \multicolumn{2}{c}{\textbf{Time (s)}} \\
         & \textbf{SF: 1K} & \textbf{SF: 1M} \\
         \midrule[1pt]
         CodexDB: - & 3 & 196 \\
         \midrule
         CodexDB: Use pandas library & 3 & 119 \\
         \midrule
         CodexDB: Use vaex library & 12 & 240 \\
         \midrule
         CodexDB: Use datatable library & 2 & 51 \\
         \midrule
         DBMS & 1 & 368 \\
    \bottomrule[1pt]
    \end{tabular}
    \label{tab:totaltimes}
\end{table}

Figure~\ref{fig:scaling} reports execution time measurements for programs generated with different instructions for the ten first queries. Missing bars indicate that no correct program was generated for the corresponding query. The data sets of the WikiSQL benchmark are too small for meaningful performance measurements. Hence, data were scaled by factor 1,000 and by factor 1,000,000 (by simply duplicating rows). The resulting data sets have an average size of 1.2~GB and 15~million rows. Table~\ref{tab:totaltimes} reports total execution time for all of the aforementioned queries for which correct programs were generated for all possible instructions. Clearly, instructing CodexDB to use different libraries has significant impact on performance. This indicates that the generated code is fundamentally different. Finally, time measurements are provided for a traditional, widely used, database management system. To ensure a fair comparison, time measurements include time for loading data from disk, processing the query, and writing the result back to disk (the generated code implements the same tasks). While performance is not the primary goal of CodexDB, the generated code is reasonably efficient.

\begin{figure}[t]
    \centering
        \begin{tikzpicture}
        \begin{groupplot}[group style={group size=1 by 1, ylabels at=edge left, xlabels at=edge bottom}, width=8cm, height=3.5cm, legend entries={-,Print ``Done.'',Print results,Print progress}, legend columns=4, ymode=normal, ymajorgrids, xlabel={Print statements}, ylabel={Nr.\ Programs}, legend to name=outputLegend, ylabel near ticks, xlabel near ticks, xtick=data, xticklabels={any, ``Done.'', string, variable}, ybar, nodes near coords, enlargelimits=0.375, every node near coord/.append style={rotate=90, anchor=west, xshift=-0.1cm}, ytick={50, 100}, ymax=110]
        \nextgroupplot[title=Davinci-Codex Model, bar width=7pt]
        \addplot coordinates {(0,2) (1,0) (2,0) (3,0)};
        \addplot coordinates {(0,100) (1,100) (2,100) (3,0)};
        \addplot coordinates {(0,100) (1,0) (2,25) (3,100)};
        \addplot coordinates {(0,100) (1,0) (2,100) (3,13)};
        \end{groupplot}
    \end{tikzpicture}
    
    \ref{outputLegend}
    \caption{Number of generated programs containing specific types of print statements for output-related instructions.}
    \label{fig:output}
\end{figure}
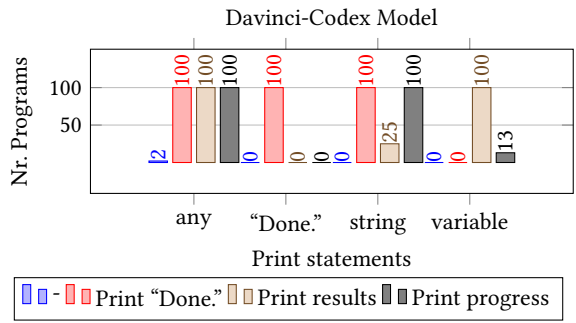

%out of 100 

Figure~\ref{fig:output} refers to logging-related instructions. The figure shows how many out of 100 generated programs contain certain types of print commands, distinguished by the operand. The figure considers presence of any print commands, commands printing out ``Done.'', commands printing hard-coded strings, and commands printing out variables. Without further instructions, only 2\% of generated programs contain any print statements. This ratio increases to 100\% for any of the logging-related instructions. Instructing CodexDB to print ``Done.'' after each step is reflected by the presence of corresponding print commands in each program. Instructing CodexDB to print intermediate results ensures that each generated program prints out variables. Requiring progress updates leads to programs printing out hard-coded strings in all (100\%) and printing out variables in some (13\%) cases. Note that this instruction leaves room for interpretation (as the form of progress updates is not specified). Manual inspection reveals that most generated code includes print commands after each step, outlining the action performed at a high level of abstraction. Figure~\ref{fig:examplegenerated} from Section~\ref{sec:overview} shows a corresponding example.
\section{Conclusion and Outlook}
\label{sec:future}

CodexDB blurs the line between user and developer. It enables far-ranging customization via natural language commands. Experiments with a first prototype are promising but also hint at significant potential for improvements. 

First, CodexDB generates correct code in most but not in all cases (up to 81\% of queries are solved, depending on scenario and model). A success rate of 100\% is illusory for any kind of natural language interfaces. Still, the newest generation of text-to-SQL methods achieves a precision of more than 90\% on the same benchmark. Hence, increasing the precision of CodexDB will be a primary research goal in the near term. %At the time of writing, OpenAI offers fine-tuning for the standard GPT-3 model but not yet for the Codex variant. Given the similarity of models, it is likely that fine-tuning Codex will soon be possible. This could yield significant gains in accuracy. Another option is to interleave code generation with code execution (allowing to integrate information gained at run time, e.g.\ on properties of intermediate result tables, into prompts). 

Second, customizing code via natural language instructions works but sometimes in unexpected ways. For instance, given instructions to use specific libraries, CodexDB always imports (``uses'') them indeed. However, in a minority of cases, imported libraries are not ultimately ``used'' for processing query steps. This motivates stronger mechanisms allowing users to enforce a specific interpretation of their natural language input. %For advanced users, this may involve inspection of generated sample code as part of the training process. Here, CodexDB is used similarly to GitHub's copilot (i.e., replacing the need to write code by the need to inspect generated code, thereby saving time for experienced developers) but with a focus on SQL processing. Enforcing a specific interpretation may be more difficult for novice users. One possible approach exploits GPT-3's capability of explaining code in natural language. Allowing users to verify whether an auto-generated code description matches their expectations may resolve some cases of misinterpretation. In other cases, users may be able to verify indirectly whether generated code matches their expectations. For instance, users can validate that output generated during query processing matches the desired format or measure the performance of generated code.

The natural language query planner does not yet use cost-based optimization. This is acceptable for the simple queries of the WikiSQL benchmark. To handle complex queries with many joins, future versions will integrate optimization according to cost models (e.g., based on the number of tuples processed) that have been shown to work quite well across different physical operator implementations~\cite{Gubichev2015}. Alternatively, machine learning can be used to optimize generated query plans for specific workloads~\cite{Marcus2018a, skinnerDB}. 

%The time required for code generation is problematic for short running queries. In the long term, CodexDB may integrate mechanisms to generate code once that is shared across queries (as opposed to generating all code from scratch for each query). Alternatively, a hybrid approach that adapts the overhead of code generation to query properties (e.g., selecting less powerful code generation models for queries on small data sets) can be used. CodexDB currently generates Python code (which Codex is most proficient in~\cite{OpenAI2021}). Generating code in different languages only requires small changes to the prompt and is planned for future versions. 

%Hence, future versions of CodexDB will support other languages such as C++ as well.

%While Codex is most proficient in Python, according to OpenAI, it supports other programming languages as well. A near-term goal is to obtain experimental results for other languages beyond Python. In the long term, specialized models for other languages may become available.
%\input{sections/conclusion}

%\clearpage

\bibliographystyle{ACM-Reference-Format}
\balance
\bibliography{library}

\end{document}